# SYNERGISM IN RADIATION EFFECTS IN CONDENSED MATTER: FUNDAMENTAL AND APPLICATION ASPECTS

Boris Oksengendler[1]*, Muhsin Ashurov[2], Sultan Suleymanov[1], Nigora Turayeva[1], Farida Iskandarova[3], Gulnoza Akhmatova[1], Rahmatillo Ibrohimov[1].

[1]Institute of Materials Science of the Academy of Sciences of the Republic of Uzbekistan, Parkent (Tashkent region), Uzbekistan
[2]Institute of Nuclear Physics of the Academy of Sciences of the Republic of Uzbekistan, "Ulugbek" settlement, Tashkent, Uzbekistan
[3]Nanotechnology Development Center at the National University of Uzbekistan named after M. Ulugbek, Tashkent, Uzbekistan
* e-mail: oksengendlerbl@yandex.ru

**Abstract**
The impressive success achieved by condensed matter radiation physics over its 170-year development period is related to the solution of problems in three areas: the emergence of new materials, the development of new sources of radiation, and the formulation of new concepts with a wide range of applications. In the borderlands of the 20th and 21st centuries, significant changes occurred in each of these aspects ("not-so-catastrophic disasters" - R. Tom). A major role was played by the emergence of a new ideology - "Complexity", which led to the birth of four paradigms; self–organization, dynamic chaos, self-organized criticality and nonadditivity (the first three are related to the concept of synergetics, the fourth - synergism).

This article is devoted to radiation synergism, with an emphasis on combinations of 2x-3x-4x radiation and other effects. It presents methods of graphical techniques to identify the specific issue of synergism effects, which is the non-additivity of the overall radiation effect. A parameter expression is proposed to account for the non-additivity of the radiation effect in the experiment, which can be compared to the q parameter introduced by Tsallis in the general science of "Complexity," opening up new possibilities in condensed matter radiation physics for both living and non-living objects.

**Keywords:** "Complexity", complex systems, synergism, diagram technique, radiation degradation, nonadditivity, Tsallis's q-entropy.

INTRODUCTION
Over the past 170 years, the initial effects of radiation exposure on media [1] have evolved into radiation physics of condensed matter (RPCM), becoming a powerful branch of science and technology, and a technology that largely defines the modern level of achievement of Civilization. It has now become evident that all leaps of progress in the field of radiation physics and technology of inanimate and animate Nature are linked to three factors: the emergence of new materials; the appearance of new irradiation equipment; and the presence (development) of new powerful theoretical concepts of radiation interaction with matter, based on the most fundamental properties of Nature [2].

The concept of **"Complexity"**, based on and developed in the form of synergetic methods, implemented in three paradigms, and synergistics, implemented in one paradigm [3,4], should be included among such concepts.
In this scheme (Fig. 1), the mechanisms leading to the birth of the new, which has become known as the special term **"emergence",** are of particular interest. But are the transitions from "normal" regimes to "synergetic" ones (studied in great detail) the only path to emergence? Apparently, one can propose other diverse variants of the appearance of the **"new,"** in particular if the "result" depends on the joint action of **several "causes"** of

**different nature**. It can be assumed (see [5]) that such a dependence, due to some special mechanisms and not leading to **synergetics**, is also realized differently. Noting that there is another "different path" to the emergence of a new quality, we can conclude that this appearance is due to the combined action of heterogeneous causes, which in radiation oncology and pharmacology is commonly referred to as synergism, implemented in scientific methodology as "synergistics." There is no doubt that such synergism (synergistics) can also be carried out in the general physics of condensed media, but in a much deeper physical formulation; related to the identification of mechanisms that can lead to synergetic effects, but not necessarily, although emergence is present here as well. In this regard, it is logical to bring both synergistics and synergetics under a single "umbrella" of Complexity - the physical meaning of such a "legacy" is due to the fact that the new "synergetic" can only be associated with nonlinearity (violation of additivity), while "synergetics" appears only under the condition of combining **nonlinearity** and **strong non-equilibrium** - Fig. 1.

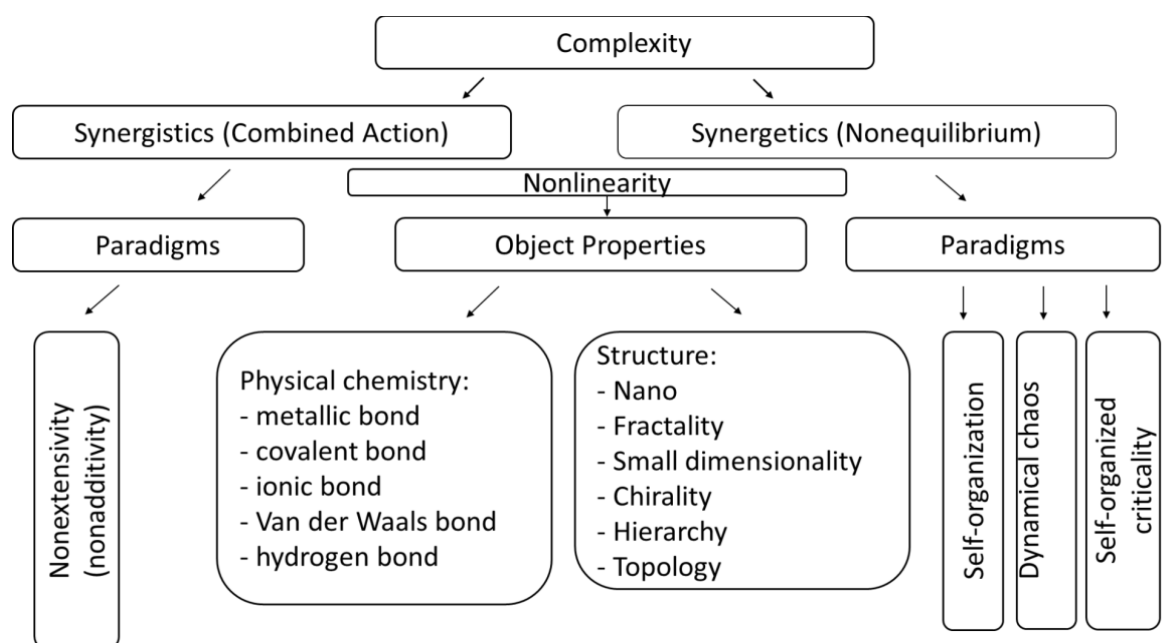

**Fig. 1.** Scheme of the expanded concept of "Complexity" in relation to radiation effects in complex media, reflecting the state of interest of researchers in the first half of the 20th century (the role of physical chemistry) and the current interest of the late 20th and early 21st centuries (the problem of structure) [6].

Let's summarize the essence of what has been said. The basic idea of synergistics (or more generally – **Synergism**) is to find the **mechanisms of non-additive enhancement** (or weakening) of the irradiation result with several impact factors. This program has been thoroughly worked out in relation to oncology, for two independent impact factors: ionizing radiation and temperature [3]. However, in general radiation physics and technology, there may be more complex cases when the number of impact factors is greater than two (see [2]). How much more complicated does the analysis of the second case become compared to the first, and what new opportunities are opened up?

## FUNDAMENTAL ASPECTS OF SYNERGISM IDEAS IN CONDENSED MEDIA

The philosophical foundations of synergism have been quite thoroughly worked out to date and are associated with so-called **holism**, as opposed to **reductionism** [5]. In the most general case, holism, in application to synergism, is represented by processes in which the

**"resulting whole"** turns out to be greater than a **simple sum**. Many situations fall under this definition, including those involving radiation exposure.

Basic Models of Synergism Phenomena with Several Impact Factors

**Case of Two Impact Factors.** Based on the above, the basic position of the concept of synergism (synergistics) in the radiation macro effect can be demonstrated in Fig. 2.

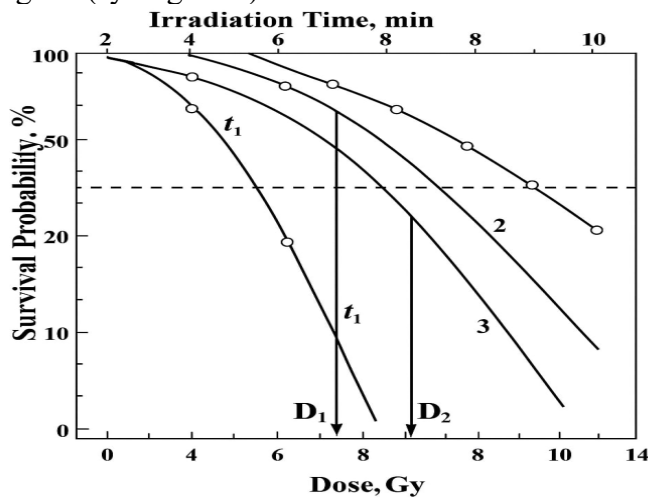

**Fig. 2.** An example of quantifying the synergistic amplification factor (an example from radiation oncology) [7].

Here it is shown how two different impact factors (ionization and external heating of the object) achieve the destruction of cancer cells [7]. It is evident that cancer cells are destroyed both by separate exposure (each of the factors lines 1 and 2), and how their combined exposure should look like with simple additivity (line 3) - a simple algebraic sum of the separate effects. However, in reality, the effect: additivity disappears, and a nonlinear effect of amplification of the combined action manifests itself in the experiment (line 4). An indicator of this amplification can be the ratio of times of equal result for the action of the theoretical and experimental curves:

$$K_2 = \frac{\tau_{th}}{\tau_{ex}} \left( = \frac{7.5}{3.8} \sim 1.9 \right) \quad (1)$$

Evidently, there is a case of amplification of the effect, meaning positive synergism. What is the mechanism of this amplification? The answer to this question can be obtained using the following algorithm: by putting forward some suitable hypothesis that allows it to be analyzed within the framework of kinetic equations [7]. Upon obtaining reasonable values for the coefficients of the kinetic equations, giving both kinetics and an amplification coefficient, it can be considered that these equations (this hypothesis) and their analysis may become a candidate, competing with the use of other hypotheses and their analysis within the framework of kinetic equations. How to select the correct hypothesis in this case? The answer to this question was obtained (within the framework of topological catastrophe theory [8]) in [6].

Returning to Fig. 2, we can formulate a hypothesis about the reasons why K is equal to ≈ 1,9 in this case, that is, greater than one. This option is shown in Fig. 3, and it consists of the following.

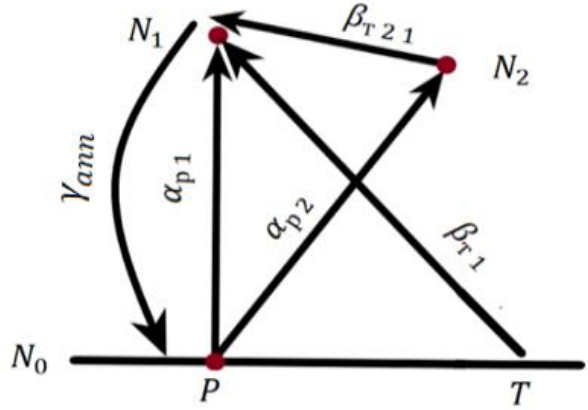

**Fig. 3.** Scheme of two-factor damage to a cancerous tumor ($\alpha_p$ – – radiation channels, $\beta_T$ – thermal channels, $\gamma_{ann}$ – annealing, $N_1, N_2$ – the number of cells converted to a state of complete and incomplete destruction, respectively.

It is visible that the initial state of the tumor ($N_0$) can transition to $N_1$, which corresponds precisely to damage. $N_1$ damage can occur in two ways: radiation *(R)* and thermal *(T)*; this corresponds to the probability $\alpha_{p\,1}$ and the probability $\beta_{T\,2}$. In addition, incomplete radiation damage to cancer cells is also possible, which corresponds to state 2. The corresponding radiation probability of this transition will be $\alpha_{p\,2}$. However, from a state of incomplete damage 2, thermal repair of the state $N_1$ is possible; it occurs with probability $\beta_{T\,2\,1}$. At the same time, the return of some fraction of cells from state $N_1$ back to state $N_0$ (return - "annealing") is not excluded; the probability of this return is $\gamma_{отт}$.

The following system of kinetic equations corresponds to the indicated scheme:

$$\begin{cases} \frac{dN_1}{dt} = N_0[\alpha_{P1} + \beta_{T1}] + N_2[\beta_{T21}] - N_1 \cdot [\gamma_{от}] \\ \frac{dN_2}{dt} = N_0[\alpha_{p2}] - \beta_{T\,2\,1}N_2 \end{cases} \quad (2)$$

The stationary solution of these equations will have the form

$$N_1 = N_0 \left[\frac{\alpha_{p2}}{\gamma_{от}} + \frac{\alpha_{P1}}{\gamma_{от}} + \frac{\beta_{T21}}{\gamma_{от}}\right] \quad (3)$$

It is visible that a number of variables depend both on temperature (through the values $\beta_T$), and on radiation parameters ($\alpha_p$); in addition, the role of the annealing parameter ($\gamma_{от}$) is important, which, in principle, can depend both on temperature and on radiation indicators.

From formulas (2), one can obtain kinetic dependencies and not only a stationary solution as in formula (3). The strategy is as follows: first we obtain the kinetics for $N_2$ (from the last expression in (2)), after which, substituting the upper expression into (2), we obtain a first-order differential equation for $N_1$.

The complete solution for the kinetics of $N_1$ will have the form

$$N_1(t) = \phi[\alpha_{r\,1}, \alpha_{r\,2}, \beta_{T\,1}, \beta_{T\,2}, \gamma_{от}; N_0] \quad (4)$$

It is important to note that all terms in the right part of this expression of the types $[\alpha_r], \gamma_{от}, \gamma_{от}^*$, depend on the intensity of irradiation, therefore all experiments where studies are focused on the role of radiation intensity are very important [10]. In particular, with $\frac{dN_2}{dt} = 0$, we get:

$$X = A[1 - \exp(-\tau)] \quad (5)$$

were $A = N_0[\alpha_{p1} + \alpha_{p2} + \beta_{T1}]$; $X = N_1 \cdot \gamma_{от}$; $\tau = \gamma_{от}t$

The found result (5) easily allows you to find the difference in kinetics for the theoretical (additive) expression and the experimentally obtained kinetics, that is $\left(\frac{dN_1}{dt}\right)_{експ} - \left(\frac{dN_1}{dt}\right)_{теор}$. A specific comparison of the obtained difference in expressions with the corresponding difference of curves 3 and 4 in Fig. 2, makes it possible to obtain complete information about all coefficients of kinetic equations.

**Case of Three Impact Factors.** In this case, we see the existence of three lines (1, 2, 3), which correspond to three "causes" - impact factors (Fig. 4).

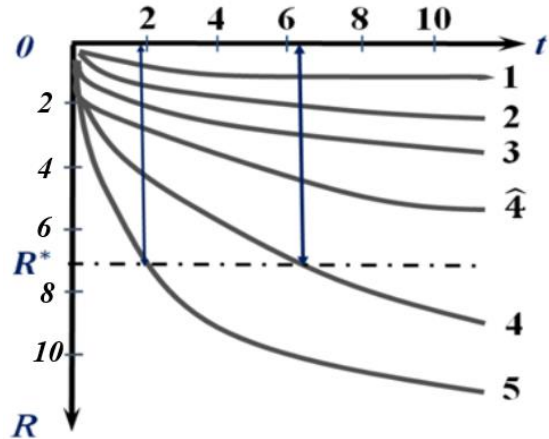

**Fig. 4.** Kinetics of the result ($R(t)$) depending on 3 factors of influence (1, 2, 3): curve 4 - in case of full synergism; curve $\hat{4}$ - in case of incomplete antagonism.

Line 4 corresponds to the simple **algebraic** sum of three experimental curves with separately acting causes. Curve 5, on the other hand, corresponds to the true "simultaneous" experiment. Pragmatically speaking, if we again draw a horizontal line of "equal effect" (the dashed line in Figure 3), then we can again write down the amplification (or weakening) coefficient depending on whether the synergism is either completely positive or (at least partially) antagonistic (which is marked in Figure 3 by curve $\hat{4}$). A very important result is evident: to achieve the $R^*$ effect, calculating the value of $K$ is not possible at all, whereas for complete synergism (line 4) it is again obtained that

$$K_3 = \frac{\tau_{th}}{\tau_{ex}} > 1; \left(\approx \frac{6}{2} = 3\right) \quad (6)$$

The interpretation of the indicated feature in Figure 4 can be built on the idea that one of the impact factors (1, 2, 3) in the case of curve 4 hinders the manifestation of the other two factors. How can you understand - many options arise here: "who hinders whom and who enhances whom"? Unlike curve 4, when all factors help each other, which gives complete additivity, the previous variant 4 is not so simple. Identifying the detailed mechanism of this situation is again possible within the framework of the method of kinetic equations according to a certain selected hypothesis. By analogy with Fig. 3, let's choose at least some (trial) scheme of combining influences, two of which (1) and (2) are ionizing radiation (such as X-ray) and external - high-temperature heating. As the third factor, let's choose, for example, an ultrasonic wave, which, being characterized by a pair of specific amplitudes and frequencies, can also destroy the neoplasm, but with a different selection of this pair of characteristics, the action of the ultrasonic wave can eliminate the metastable states formed as a result of the primary "breakdown" of molecules. Such a possible scheme of three influences is shown in Fig. 5.

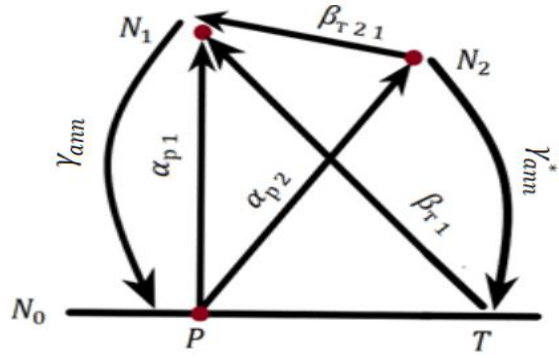

**Fig. 5.** Scheme of three-factor effect on a cancerous tumor (compared to Fig. 3. here, the decay of the $N_2$ state is added with return to the initial state $N_0$; the probability of this process is described by the expression $\gamma^*_{ann}$.

The following system of kinetic equations corresponds to the indicated scheme:
$$\begin{cases} \dfrac{dN_1}{dt} = N_0[\alpha_{P1} + \beta_{T1}] + N_2[\beta_{T21}] - N_1 \cdot [\gamma_{oT}] \\ \dfrac{dN_2}{dt} = N_0[\alpha_{p2}] - (\beta_{T21} + \gamma^*_{oT})N_2 \end{cases} \quad (7)$$

The stationary solution of these equations will have the form
$$N_1 = N_0\left[\frac{\alpha_{P1}+\beta_{T1}}{\gamma_{oT}} + \frac{\beta_{T21}\cdot\alpha_{P2}}{\gamma_{oT}(\beta_{T21}+\gamma^*_{oT})}\right] \quad (8)$$

This nonlinear by $\gamma^*_{oT}$ expression, with a weak channel for eliminating the metastable state $N_2$, gives a linear decrease (by $\gamma^*_{oT}$) in the output of the effect $N_1$, which can translate curve 4 to curve $\hat{4}$ (see Fig. 4).

It is essential that the system of equations (7) allows a general solution of kinetics (and not just a stationary variant), in particular, similarly to how it was performed before obtaining formula (5), in this case, provided that it is stationary only for the second equation ($\frac{dN_2}{dt} = 0$), we get

$$X^* = A^*[1 - \exp(-\tau)] \quad (9)$$

Where $A^* = N_0\left[\alpha_{p1} + \alpha_{p2}\left(1 + \frac{\gamma^*_{oT}}{\beta_{T21}}\right)^{-1} + \beta_{T1}\right]$; $X^* = N_1 \cdot \gamma_{oT}$; $\tau = \gamma_{oT}t$

A completely different type of realization of the third impact factor is also of interest. Thus, illumination in the region of the visible spectrum of waves, as is known [6], leads to the recharging of certain local electronic states of reorganizing chemical bonds, which affects their stability (or instability) and is responsible for the occurrence of chemical processes upon irradiation with X-rays with additional illumination. Perhaps, it is such effects that are responsible for the peculiarities of the destruction processes found in a number of radiobiological works (see [10]).

Alongside the creation or elimination of metastable states, in principle, there is a completely different type of nonlinearity implementation when an unusual impact factor is attracted to the two factors previously involved. In the event that electrons from outer shells (valence electrons) are involved in the physicochemical-biological processes, the role of spin degrees of freedom in processes influencing chemical reactions may be revealed. We are talking about the so-called spin chemistry [11] in complex systems, when the implementation of the process is associated with transitions of the type singlet → triplet state and vice versa, in particular, when absorbing photons. The physical essence of such processes is as follows: before S→T (singlet to triplet in a two-electron covalent bond), however, in order for such a transition to occur and cause local destruction of the covalent bond, it is necessary, in addition to the participation of a UV photon, also to satisfy the Wigner symmetry rules, which requires the involvement of a third spin in the process of flipping one of the spins of the chemical bond (flip-flop process). When the appropriate conditions are met, all stages of molecular reorganization occurring at the chemical stage of the radiation process can radically change, even by involving nuclear spins. This spin catalysis, which is an indisputable fact of modern magnetic photochemistry, can be used completely radically, and, if necessary, controllably, to increase either the synergism or the antagonism of all factors of influence of the general macro effect; as far as we know, in radiation oncology, where two-factor synergism has been studied in great detail [10]), the idea of spin catalysis was not put forward or used, although the magnitudes of magnetic fields suitable for such control are very small [11-12].

**Case of Four Impact Factors.** But how should we proceed in the case of many causes – impact factors? You can use the same technique as in Fig. 1, followed by analysis using kinetic equations that take into account four types of exposure; obviously, such a solution "head-on" is very difficult. In this situation, it is interesting to use a different strategy: to trace (experimentally and theoretically) the synergism of six possible pairwise combinations of all 4 factors, and on this basis calculate the partial coefficients of synergism ($K_{12}, K_{13}, K_{14}, K_{23}, K_{24}, K_{34}$) and, using which, to estimate the total coefficient of synergism, taking into account the simultaneous impact of 4 components, as well as to test this new algorithm on a suitable object. Such an object, which, in our opinion, may be the so-called Soret effect [13-14], which is the ionic electron transfer in a condensed medium under the influence of a "phonon wind" caused by a temperature gradient, on superionic materials (such as $LaF_3$, $CeF_3$). The idea of the experiment is as follows: by changing (increasing) the temperature of the experiment, it is possible to achieve a phase transition from a non-superionic state to a superionic one; at the same time, the value of the Soret coefficient increases sharply - after the appearance of the superionic phase; the ionic transfer effect itself is commensurate with the temperature gradient, as well as the presence of an ionizing component of solar radiation (at the Large Solar Furnace, LSF); (the latter circumstance implements the so-called electron-stimulated radiation diffusion [9]). Thus, the pair **"temperature/temperature gradient"** is easily singled out. By varying the experimental conditions, it is possible to single out (as dominant) other pairs of impact factors of the radiation field of the LSF.

Separately, it should be mentioned the possibility of fluctuations in the density of ions due to the formation of spatially-resolved regions resulting from the superionic transition. Such local areas with altered density - easily movable quasi-liquids of ions, are interestingly reminiscent of the so-called negative crowdions of Indenbom-Kratochvill [15], which are in the field of action of the temperature gradient (in particular, the capture of electrons from the conduction band to local states that stimulate the movement of crowdions is especially interesting here [8]. A significant (and still unresolved) question is how the phonon field should be described in an object where strictly periodic and disordered regions are systematically present and adjacent. From a practical point of view, the possibility of the Soret effect to become an indicator of the presence (or absence) of precisely the electronic component as a result of irradiation of the quasi-liquid with the UV - part of the spectrum is very curious.

From the point of view of the strict theory of irreversible thermodynamics, the case of the "phonon wind" should be attributed to the so-called thermodiffusion [16], in which the **Soret parameter** ($D^T/D$) arises, which connects the gradients of concentrations of mobile ions and temperature, that is

$$\frac{gard\ C}{grad\ T} = \frac{CD^T}{D} \qquad (10)$$

Here $C$ is the concentration of the transported ions, $D$ is the usual diffusion coefficient, $D^T$ is the thermodiffusion coefficient. A fundamentally new element - the Soret coefficient, and it has been repeatedly discussed at various conferences on ionic transfer [13-14,16]; here we will limit ourselves to the "semiconductor" approaches of Fix [13] to calculating the force, thanks to which non-equilibrium phonons (temperature gradient!) when scattered on impurities or inhomogeneities "drive" them against the temperature gradient, forcing them to accumulate in areas with reduced temperature of the medium. Fix's work shows that the flux of particles carried away by the force of the "phonon wind" $F_{ϕi}$,

$$J_T = C_{iui}F_{ϕi} = J_T - C_I S_{ϕi} D_i \frac{dT}{dx} = -C_i \frac{α_{ϕi}}{T} D_i \frac{dT}{dx} \quad (11)$$

Here $S_{ϕi}$ is the Soret coefficient corresponding to is equal to the entrainment of the impurity, $α_{ϕi}$ is the thermodiffusion coefficient, $D_i$ is the diffusion coefficient of the impurity. The Einstein approximation gives for the Soret coefficient:

$$S_{ϕi} = -\frac{F_{ϕi}}{kT}\left(\frac{dT}{dx}\right)^{-1} \quad (12)$$

Further, the task is reduced to calculating the estimate of the force $F_{ϕi}$, for which it is necessary to move away from the general approach and be satisfied with suitable models of the scattering center and the phonon field, as well as a model of the details of their interaction. Fix showed [13] that

$$F_{ϕi} = -\frac{1}{3} C_V l^-_{α_{ϕi}} \frac{dT}{dx} \quad (13)$$

$$S_{ϕi} = \frac{1}{3} \frac{C_V l^-_{α_{ϕi}}}{kT} \quad (14)$$

Comparing formulas (10-13), we see that the magnitude of the ion flux, which can be measured, easily varies both with temperature and with its gradient (in the LSF).

## SOME APPLIED ASPECTS OF SYNERGISM

The fundamental features of combined radiation interaction with matter, discussed above, as shown in the model case of "ionizing radiation + heat", allows for a quite adequate analysis within the framework of kinetic equations: this makes it possible in some cases to identify the effects of synergism, transgressive factors, which leads to the loss of additivity of the irradiated object. Note that the effects of synergism in radio-oncology have gradually begun to penetrate into other radiation-technological procedures, which have both much in common and many differences for the phenomena of animate and inanimate nature.

Next, we will discuss some successes in this area, which are still weakly mathematized.

### Thermo-radiation effects in complex non-metallic solids

Many years of research in the radiation physics of nonmetals have long raised the question: how does radiation exposure differ from thermal exposure (see [17-20]). Indeed, there is the so-called radio-biological paradox: A radiation dose of more than five grays is sufficient for irreversible (lethal) damage to a biological system (e.g., the tobacco mosaic virus), but if an equivalent amount of energy is used to heat a glass of water, there will be no tragic destruction of this system. Now the answer to this question is, in fact, trivial: in most cases, ionizing radiation damages the molecules of the medium locally, and thermal exposure - non-locally. From this it is clear that a rich selection of the desired type of irradiating radiation is very interesting precisely in combination with heat (own or external) to control the required output of the radiation effect. This is quite solvable when using the Platzman-Starodubtsev scheme [2, 20] to describe radiation exposure: firstly, from taking into account that any types of radiation exposure at the elementary level can be described within only five channels of energy transfer from radiation to matter (ionization, elastic scattering, heat, thermal and shock waves [3]). However, to date, this program of combining knowledge on the multistage nature of radiation effects and all five channels of energy transfer (at each stage!) has not been solved, although preliminary successes already exist [2,3,20]. But a completely new chapter in the synergy of the emergence of nonlinearity opens up in such irradiation objects as the synchrotron Large Solar Furnace and the pulsed emission electron accelerator [3].

Synergism in Degradation Problems of Certain Laser Materials

The degradation of complex materials and devices based on them is one of the most critical problems in modern solid-state electronics and many other structures [18, 21-22]. Efforts to combat degradation (particularly in semiconductor devices and lasers) have shown that as materials and device structures become more complex, the fight against degradation becomes significantly more challenging [18, 21-22]. A characteristic example of this is the long-standing and not very successful struggle to increase the efficiency of perovskite-based solar cells. From our perspective, this situation is far from coincidental: the limited success in suppressing degradation in such complex systems requires more sophisticated methods based on a modern understanding of complex systems in general [23].

A crucial point is the consideration of the system's openness, in which device connections to the external world are realized through flows of matter, energy, and information [24]. In this context, the extraordinary similarity between the general functioning and failure processes of complex non-living systems (devices) and various objects in living Nature becomes apparent. It can be assumed that in both cases, the normal operation of an object is achieved only with a specific balance of order and disorder in their operating modes, deviation from which constitutes "degradation" or "disease."

In this case, the optimal operating (functioning) condition of a complex system is the *intermittency regime* [25], where accumulating harmful deviations (during the normal stage) are eliminated ("washed away") in chaotic regimes, after which a phase of normal operation resumes with the previous balance of order and chaos [21] Fig. 6.

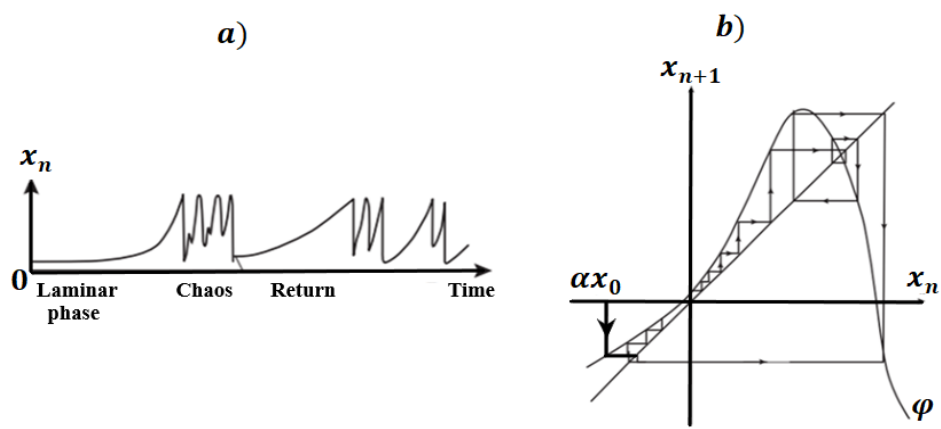

**Fig. 6.** Functioning in the interleaving mode of a complex object (instrument) under conditions of radiation exposure: a) – combination of sequential "laminar" and "turbulent" stages; b) – modeling of the "loop" return process of a complex object based on the Lamerey diagram [21-26].

Here's the translation of the next section:

It seems to us that such a picture of normal device functioning is particularly optimal under conditions of radiation exposure. Naturally, such a "looping" mode of device operation can be ensured with a deep understanding of the physics of radiation-induced atomic rearrangements, their organization into a closed loop, taking into account the general radiation scheme of Starodubtsev-Platzman [3, 20] and the special property of complex systems – their *integrity* [23], which in the specific case of O-O bridge destruction is achieved by shifting the quasi-Fermi level ("down"), which ionizes electrons of paramagnetic ions and inhibits *f-f* processes [6]. Control of the parameters of the resulting looping mode is carried out based on the laws of cybernetics [27].

Taking into account the combination of external influence (radiation) with the special properties of the object (integrity) is especially evident in structures built on the laws of hierarchy. In our case of radiation exposure on complex solid-state structures, we should talk about the separate impact of radiation on electronic and atomic subsystems, connected by special kinetic regularities that implement forward and reverse connections. In the search for the diversity of such modes, it is very convenient to use the ideas of synergism and synergetics, which reveal special mechanisms for amplifying or suppressing deviations from simple additivity of effects [23]. Note that the above is very clearly described within the framework of the Lamerey diagram (Fig. 6); such a scheme was implemented, e.g., in [28].

An important example of the above are solid-state electronics materials containing so-called "oxygen bridges" *O-O* (Fig. 7), often components of glasses used in lasers and other devices. The peculiarity of these bridges is that both oxygens are connected to each other only by a covalent bond. A large number of studies on the radiation impact on similar objects (see [18]) lead to the conclusion that *O-O* bridges are the weak link, the accumulation of defects of which during operation leads to the degradation of laser parameters. Taking into account the emergence of spin catalysis ideas, these general considerations can be specified. Indeed, according to the scheme presented in Fig. 7, it can be assumed that radiation exposure (e.g., X-ray radiation or *UV*) can stimulate a *flip-flop* type process, converting a singlet covalent bond into a triplet, naturally, with the participation of a third-party spin, belonging either to a nearby paramagnetic ion or to a fast electron born in ionization processes by radiation.

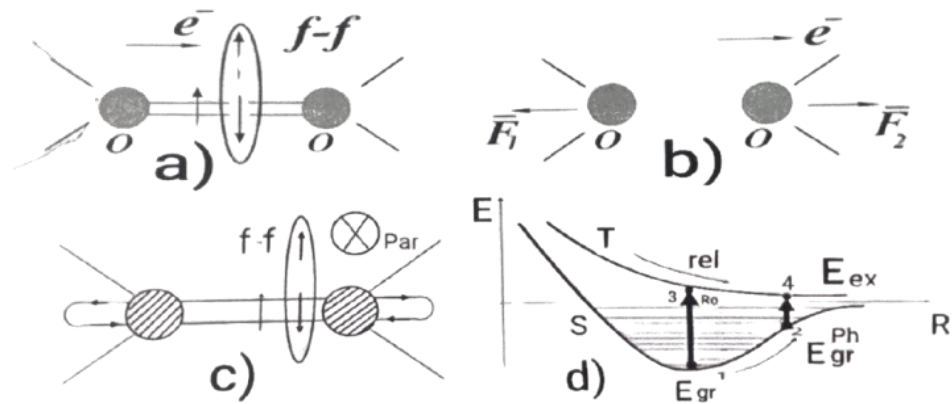

**Fig. 7.** Scheme of the *O–O* bridge decay by the *flip-flop* mechanism under the influence of X-ray or *UV* radiation: *a* – the initial position with the S state of the oxygen bridge electrons; *b* – the state of the oxygen bridge after the *f-f* process leading to the *T* state of the oxygen bridge electron pair; *c* – the decay of the oxygen bridge after the S→T transition; *d*) – the scheme of the *O–O* bridge decay as a result of the virtual *f-f* process upon photon absorption [22, 31].

Here, it is important to note that the formed triplet, repelling oxygen ions, does not always lead to a defect, but only in those cases when the electronic level of the triplet state, being in the conduction band, becomes resonant. This leads to the possibility of its detachment from the local state and movement through the free band, so that the oxygen ions do not have time to diverge before that. In this case, the probability of destruction of the *O-O* bridge is multiplied by a factor $\eta = \exp(-\tau_+/\tau_e) \approx \exp(-\Delta E_r/\hbar\omega_D)$; here $\Delta E_r$ is the width of the resonant level, $\omega_D \approx 1/\tau_+$ is the Debye frequency [29-30]. If we consider that there are many structures of the *O-O* bridge type, then such a mechanism can be responsible for degradation in a large number of situations (e.g., in a number of HTSCs and third-generation solar cells). It is essential that the f-f process can also be useful in some radiation technology processes (e.g., under the conditions of operation of the Large Solar Furnace [3]). All this makes the *f-f* mechanism a contender for the "title" of universal, provided that there are various bridge structures in the irradiated object.

Synergism in Modern Approaches to Radiation Inactivation of Viruses

The dramatic events that followed as a result of the Covid-pneumonia pandemic caused by the *SARS-2V* type coronavirus, which began in December 2019, revealed a very weak effectiveness of pharmacology in combating this disease and its causes [17]. In this situation, radiation physics played a very special role – irradiation with low-energy *X*-rays, and in small doses. As a micro-mechanism of *X*-ray inactivation, the mechanism of Auger destruction of the RNA of the *SARS-2V* virus was proposed, and taking into account the peculiarities of its structure revealed a greater efficiency of destruction of the corresponding *RNA* loci (in vivo) compared to the characteristic probabilities of destruction of ordinary biopolymers in cells. It was precisely the probability of increasing the Auger destruction of the *RNA* virus by approximately $10^3$ times, compared to other biopolymers, that predicted the success of the low-dose therapy method, when small doses of irradiation are sufficient to destroy viral *RNA*, while for ordinary biopolymers of the environment, these doses are of little significance [20]. A detailed analysis of the Auger destruction of *SARS-2V RNA* showed [17] that medical results on low-dose irradiation of Covid-pneumonia, published shortly after theoretical studies [32], turned out to be quite effective (see [33]) in accordance with the prediction and experiment. At the same time, even small doses of *X*-ray radiation (less than 1 Gray) still pose a potential danger to the body: as a result of each Auger transition, from which the general Auger cascade is constructed, an Auger electron is emitted, having an energy of the order of hundreds of electron volts. Such electrons with a noticeable probability excite chemical bonds of nucleotide bases, which, in turn, are precursors of mutations (Fig. 8):

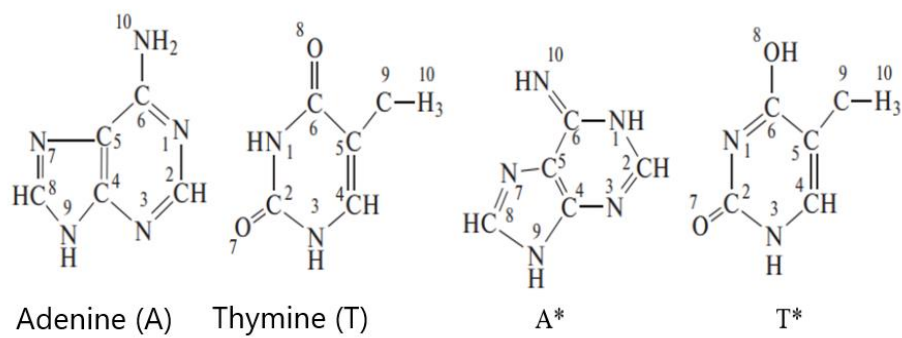

**Fig. 8.** Simplified scheme of premutation formation due to electron-stimulated hydrogen transitions: 10→1 (in adenine) and 1→8 (in thymine) [34].

As a result, we get a rather unpleasant combination of two events: first, the Auger cascade (e.g., in a phosphorus atom), reaching the valence shells of the incident phosphorus atom and its nearest neighbors, performs a "Coulomb explosion" in the local region of *DNA* [17] - this fact has been found in a number of works [35], so that the selected *RNA* locus *SARS-*

$2V$ is destroyed; on the other hand, these "accompanying" Auger electrons are generators of premutations, which effectively lead to oncological mutations [36].

It is interesting to consider whether nature has provided any mechanism of synergism that resolves this duality, namely: beneficially destroys the virus locus and beneficially eliminates premutations?

Apparently, such a synergism option does exist, and this is what it boils down to. Referring again to Figure 8, it is necessary to say that all states with an asterisk $A^*$ and $T^*$ are metastable with respect to $A$ and $T$, respectively. This circumstance is reflected in Fig. 9 [17]:

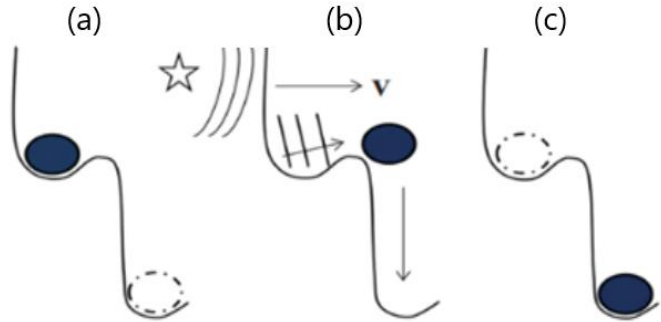

**Fig. 9.** Configurations of the hydrogen atom: *(a)* – metastable (tautomeric form); *(b)* – intermediate ("squeezing" of the hydrogen atom from the metastable state to the stable state under the action of a shock wave generated by a Coulomb explosion); *(c)* – returned ground state (stable configuration).

Obviously, the most important property of a successful discharge of a metastable configuration into a stable one ($a→б→в$) is related to the fact that the speed of the incoming shock wave $(\vec{V})$ is greater than the average speed of the oscillating atoms ($|\vec{v}_a|$): then the depth of the metastable state will be eliminated faster than the "bump" in the metastable state will shift, running away from the shock wave, $|\vec{V}| > |\vec{v}_a|$. It is essential that this inequality is realized with a margin precisely as a result of the dispersal of "fragments" of molecular structures during a "Coulomb explosion" initiated by the Auger cascade. Thus, the shock wave generated by the Coulomb explosion "cleans" the environment near the incident Auger cascade from metastable states formed by accompanying Auger electrons.

Simple estimates based on the theory of elasticity allow us to estimate the volume of the cleared region.

$$\Omega \approx d_0^3 [E_{coul}/E_{met}] \quad (15)$$

Here, $d_0^3$ is the volume of the initial localization of the Auger charge, $E_{coul}$ and $E_{met}$ are the energies of the "Coulomb explosion" and the metastability barrier.

The estimated cleaned volume is approximately $1{,}5 \cdot 10^4$ Å$^3$.

We would like to hope that this version of synergism will prove effective for the processes of "ordinary" radiation oncology in humans [37-40].

Radiation synergism from the perspective of the Tsallis approach to complex systems.

In Figure 2, the segments $|N_4(t) - N_3(t)|$ set the level of non-additivity at different doses (times) of combined radiation and heat exposure to a cancerous tumor. Note that

$\Delta N(t) = |N_4(t) - N_3(t)|$ increases with increasing dose, which means that the positive interaction between both factors increases as the tumor is destroyed. On the other hand, as is known [41], the modern interpretation of a "complex system" requires that such a system must have a number of characteristic properties, and above all, "non-additivity" (nonextensivity), which is extremely effectively studied based on Tsallis' approach [42] to q-deformation statistics and thermodynamics. Indeed, in the theoretical concept of Tsallis [41], the non-additivity of states is described by q-deformation statistics and q-deformation thermodynamics, through which Tsallis' q-entropy is expressed:

$$S_q = -\sum_i p_i \left(\frac{P_i^{q-1} - 1}{q-1}\right) \quad (16)$$

Here, $p_i$ is the Boltzmann probability of state $i$, and the important equation $Q_p = \frac{|N_4(t) - N_3(t)|}{|N_4(t) + N_3(t)|} = q$ defines the possibility of creating a "bridge" between radiation exposure experiments (e.g., X-ray exposure, Fig. 2) and the Tsallis theory, especially since the modern version of $q$ is a $q$-triplet [41-43], which opens up fundamentally new possibilities in radiation science and practice, both for non-living and living systems.

## CONCLUSION

The history of radiation physics in objects of living and non-living Nature clearly indicates three reasons for the sharp development of this scientific field: the appearance of new materials (objects); the appearance (discovery) of new types of radiation exposure; and the appearance of new, major theoretical ideas affecting all branches of science about matter, which after a short time also spread to the study of radiation properties. And so, leaps forward in our understanding are always due to new major concepts; there have been quite a few of them in the past 170+ years since the beginning of the study of radiation properties. Without a doubt, in the last 25 years, new ideas about the structure of matter, as well as ideas about the combined action of radiation exposure factors, which could be obtained from one type of radiation or from several simultaneously, have played a special role. Both proved important in materials science, but also in radiobiology (with the transition to radiation medicine). In radiation medicine, ideas about the combined action of different radiation sources were particularly successfully developed, which were combined under the name synergism (in particular, in oncology). In radiation physics and chemistry of non-living objects, the ideas of using various factors of radiation action depending on its energy and intensity were more popular. However, recently, in connection with the clarification that modern radiation physics is a science about complex materials, these approaches have become intertwined and united under the term "Complexity," the methodology of which is divided into two approaches: synergetics and synergistics (Figure 1). Synergistics has turned out to be less developed to date, which has given rise to its deep mastering in the radiation physics of non-living objects. In this article, an attempt was made to systematically consider the models of synergistics and their application equally in both living and non-living Nature. Thus, it seems to us that we have succeeded in identifying new methodological approaches in synergistics – on the one hand, and on the other hand – in showing how classical methods of radiation physics can be usefully mastered in the concepts of synergism.

## FUNDING
This work is funded by the state budget of the Republic of Uzbekistan.

## CONFLICT OF INTEREST
The authors of this work declare that there is no conflict of interest.